\documentclass[aps,pra,showpacs,showkeys,superscriptaddress]{revtex4}
\usepackage{amsfonts,amsmath}
\usepackage[dvips]{graphics}

\begin{document}

\newcommand{\ket}[1]{\mathop{\left|#1\right>}\nolimits}       
\newcommand{\bra}[1]{\mathop{\left<#1\right|}\nolimits}       
\newcommand{\Tr}[1]{\mathop{{\mathrm{Tr}}_{#1}}}              
\newcommand{\braket}[2]{\langle #1 | #2 \rangle}
\newcommand{\ketbra}[2]{| #1\rangle\!\langle #2 |}
\newcommand{\bbR}{{\mathbb R}}
\newcommand{\bbZ}{{\mathbb Z}}
\newcommand{\ahead}[2]{\genfrac{}{}{0pt}{}{#1}{#2}}
\newcommand{\nn}{\nonumber}

\title{Continuous variable private quantum channel}
\author{Kamil Br\'adler}
\affiliation{Department of Chemical Physics and Optics, Charles
University, Ke~Karlovu 3, 121~16 Prague~2, Czech~Republic}

\email{kbradler@epot.cz}

\affiliation{Department of Optics, Palack\'y University,
17.~listopadu 50, 772\,00 Olomouc, Czech~Republic}

\date{\today}

\begin{abstract}
In this paper we introduce the concept of quantum private channel
within the continuous variables framework (CVPQC) and investigate
its properties. In terms of CVPQC we naturally define a
``maximally" mixed state in phase space together with its explicit
construction and show that for increasing number of encryption
operations (which sets the length of a shared key between Alice
and Bob) the encrypted state is arbitrarily close to the maximally
mixed state in the sense of the Hilbert-Schmidt distance. We bring
the exact solution for the distance dependence and give also a
rough estimate of the necessary number of bits of the shared
secret key (i.e. how much classical resources are needed for an
approximate encryption of a generally unknown continuous-variable
state). The definition of the CVPQC is analyzed from the Holevo
bound point of view which determines an upper bound of information
about an incoming state an eavesdropper is able to get from his
optimal measurement.
\end{abstract}



\maketitle

\section{Introduction}

The task of quantum state
encryption~\cite{quant_ver_pad,PQC,encryption} (quantum Vernam
cipher, quantum one-time pad) is defined as follows. Let's suppose
that there are two communicating parties, Alice and Bob, and Alice
wants to send an arbitrary unknown quantum state to Bob (the state
needn't to be known either Alice or Bob). They are connected
through the quantum channel which is accessible to Eve's
manipulations. To avoid any possible information leakage from the
state to Eve (via some kind of generalized measurement or state
estimation) Alice and Bob share a secret and random string
(classical key) of bits by which Alice chooses a unitary operation
from the given (and publicly known) set. The transformed state is
sent to Bob via the quantum channel who applies the same operation
to decrypt the state. For an external observer (e.g. Eve) the
state on the channel is a mixture of all possible transformations
that Alice can create because she doesn't know the secret string.
If, moreover, the mixture is independent on the state to be
encrypted then Eve has no chance to deduce any information on the
state. We say that the state was perfectly encrypted. After the
formalization of this procedure we get the definition of {\it
private quantum channel -- PQC}~\cite{PQC}. As the mixture it is
advantageous to choose a maximally mixed state -- identity density
matrix.

It is natural to ask how many operations are needed for encryption
of a $k$-qubit input state. Based on thoughts on entropy it was
shown~\cite{encryption} and later generalized~\cite{PQC} for PQC
with ancilla that $2k$ classical bits are sufficient (i.e.
$2^{2k}$ operations). Generally, for perfect encryption of
$d$-dimensional quantum state $d^{2}$~unitary operations are
needed. Thus, the length of participant's secret and random key
must have at least $2\log d$ bits~\footnote{Throughout this paper,
all logarithms have base two.}. After weakening the security
definition an approximate secrecy was stated
in~\cite{approx_encryption} defining of {\em approximate private
quantum channel -- aPQC}. Then, it was shown that asymptotically
just $d\log d$ operations are needed for approximate quantum state
encryption. Next progress in this research topic was achieved
in~\cite{approx_encryption1} where a polynomial algorithm for
constructing a set of encryption/decryption operations suitable
for aPQC was presented.

In this paper we investigate the possibility of quantum state
encryption for continuous variables (CV)~\cite{CV}. Under CV we
understand two conjugate observables such as e.g. position and
momentum of a particle. Especially, we concentrate on coherent
states states which Wigner function has the form of the normalized
Gaussian distribution. Gaussian states are the most important
class of CV states used in quantum communication and
computation~\cite{CV_prehled}. The most of intriguing processes
and algorithms discovered for discrete $d$-level quantum systems
were also generalized for~CV. Among others, let's mention CV
quantum state teleportation~\cite{CV_tele1,CV_tele2}, CV quantum
state cloning~\cite{CV_clone} and quantum computation with
CV~\cite{CV_comp}. Importantly, a great progress was made in
quantum key distribution (QKD) based on CV where theoretical
groundwork was
laid~\cite{CV_crypto1,CV_crypto2,CV_crypto3,CV_crypto4,CV_crypto5}
and experiments were
performed~\cite{CV_crypto_exp1,CV_crypto_exp2}.

After brief introducing into the questions of distances used in
quantum information theory in Sec.~\ref{sec_intro} the main part
follows in Sec.~\ref{sec_main} and Sec.~\ref{sec_threats}. There
we present the notion of CV quantum state encryption and in the
later we define a continuous-variable private quantum channel
(CVPQC). This, foremost, consists of defining the ``maximally''
mixed state within the context of continuous variables and
estimating the length of a secret shared key between Alice and Bob
for a given secrecy (by secrecy it is meant the HS~distance
between ``maximally" mixed and the investigated state). In
Sec.~\ref{sec_threats} we will touch the question of
``maximality'' of the mixed state (from now on, let's omit the
apostrophes) in the context of bosonic
channels~\cite{bos_chan1,bos_chan2} and their generalized
lossy~\cite{bos_chan_lossy} and Gaussian
relatives~\cite{bos_gauss_chan_mem}. We will also discuss
important differences between discrete and CV encryption from the
viewpoint of eavesdropping followed by the calculation of the
Holevo bound limiting information accessible to Eve from the
encrypted channel. Necessary Appendices come at the end of the
paper. In Appendix~\ref{app_exact_value} we give a derivation of
the exact formula for the HS~distance. The object of
Appendix~\ref{app_guess} is to inference the mentioned estimate of
the HS~distance.

\section{Measures of quantum states closeness}
\label{sec_intro}

Quantum states can be distinguished in the sense of their mutual
distance. The distance is usually induced by a norm defined on the
space of quantum states. This is the case of Schatten $p$-norm
\begin{equation}\label{schatten}
    \|A\|_p
    =\left(
       \Tr{}\left(\left|A\right|^p\right)
    \right)^\frac{1}{ p}
    =\left(
       \Tr{}\left(A^p\right)
    \right)^\frac{1}{ p},
\end{equation}
where $|A|=\sqrt{A^\dagger A}$ and the last equation is valid for
$A=A^\dagger$. If $A=\varrho_1-\varrho_2$ the Hermiticity of $A$
is still preserved and for $p=2$ we get the Hilbert-Schmidt~(HS)
distance
\begin{equation}\label{HSdist}
    D_{HS}(\varrho_1,\varrho_2)
    =\|\varrho_1-\varrho_2\|_2=\sqrt{\Tr{}\left((\varrho_1-\varrho_2)^2\right)}.
\end{equation}
On the other hand, there is a whole family of distances based on
Uhlmann's fidelity~\cite{uhlmann}
\begin{equation}\label{fid_uhl}
F(\rho_{1},\rho_{2})=\Tr{}\left(\left(\sqrt{\rho_{1}}\rho_{2}
\sqrt{\rho_{1}}\right)^\frac{1}{2}\right).
\end{equation}
One of them is Bures distance~\cite{bures}
\begin{equation}\label{buresdist}
    D_B(\varrho_1,\varrho_2)
    =\left(2-2\Tr{}\sqrt{\sqrt{\varrho_1}\varrho_2\sqrt{\varrho_1}}\right)^\frac{1}{2}
\end{equation}
which coincides with the HS~distance if $\varrho_1,\varrho_2$ are
pure states.

There is a certain equivocation which distance is more suitable
for a given task in QIT where a general problem of quantum state
distinguishability or closeness has to be resolved. Both distances
have many useful properties and they are subject of detailed
investigation~\cite{SommZyc}. For example, output ``quality" of a
quantum state in the problem of universal quantum-copying machine
was first analyzed with the help of the HS~distance~\cite{UQCM_HS}
and later revisited from the viewpoint of the Bures and trace
distance (Schatten $p$-norm~(\ref{schatten}) for
$p=1$)~\cite{UQCM_Bures}. Other problem, among others, where
closeness of two quantum states plays a significant role is
quantum state estimation~\cite{quantstateest}. Here the closeness
of estimated states is often measured by the
HS~distance~\cite{quantstateest_HS}.

The motivation for using the HS~distance in our calculation is
two-fold. First, the security criterion for approximate quantum
state encryption~\cite{approx_encryption} is based on the operator
distance induced by operator norm ($p\to\infty$
in~(\ref{schatten})) while some other results therein are
calculated for the trace distance which is in some sense weaker
than the operator distance (the reason is computational
difficulty). Second, for our purpose we need to calculate distance
between two infinitely-dimensional mixed states what is difficult
in the case of all fidelity based distances. However, we suppose
that the HS~distance is a good choice for coherent states and
provides an adequate view on measure of the closeness of two
states.

\section{CV state encryption and its analysis}
\label{sec_main}

Let's define the task of CV state encryption. As in the case of
discrete variable quantum state encryption, Alice and Bob are
interconnected via a quantum channel which is fully accessible to
Eve's manipulations. Both legal participants share a secret string
of random bits (key) which sufficient length is, among others,
subject of this paper. The key indexes several unitary operations
which Alice/Bob chooses to encrypt/decrypt single-mode coherent
states. The purpose of the encryption is to secure these states
from leakage of any information about them to an eavesdropper
(Eve). The way to achieve this task is to ``randomize" every
coherent state to be close as much as possible to a maximally
mixed state (maximally in the sense specified next). The
randomization is performed with several publicly known unitary
operations (it is meant that the set from which participants
choose is known but the particular sequence of operations from the
set is given by the key which is kept secret). So, suppose that
Alice generates or gets an arbitrary and generally unknown
single-mode coherent state. The only public knowledge about the
state is its appearance somewhere inside the circle of radius $b$
in phase space with the given distribution probability. Here we
suppose that states occur with the same probability for all $r\leq
b$ and with zero probability elsewhere.

Suppose for a while that Alice encrypts a vacuum state. This
situation will be immediately generalized with the help of the
HS~distance properties for an unknown coherent state within the
circle of radius $b$ as stated in the previous paragraph. The
first problem we have to tackle is the definition of a maximally
(or completely) mixed state. Here the situation is different from,
generally, $d$-dimensional discrete Hilbert space where a
normalized unit matrix is considered as the maximally mixed state.
This is inappropriate in phase space nevertheless we may inspire
ourselves in a way the discrete maximally mixed state is
generated. In fact, we get the maximally mixed state (in case of
qubits $\varrho_V=\varrho_{r\vartheta\varphi}$) by integrating out
over all density matrix populating the Bloch sphere
$\openone\propto\int\varrho_V{\rm d}V$ (irrespective of the fact
that the finite number of unitary operations suffice for this task
as the theory of PQC learns). Similarly, as a maximally mixed
state~\footnote{In fact, Eq.~(\ref{integral_0b}) belongs to the
class of bosonic channels later discussed
in~section~\ref{sec_threats}. We will address the problem of the
``measure of maximality'' of the mixed state in the context of the
calculated Holevo bound on the channel.} we can naturally choose
an integral performed over all possible single-mode states within
the circle of radius $r\leq b$
\begin{align}\label{integral_0b}
\openone_b
    & =\frac{1}{ C}\int{\rm d}^2\alpha\ketbra{\alpha}{\alpha}
     =\frac{1}{ C}\sum_{m,n=0}^\infty\ketbra{m}{n}\int_0^{2\pi}
     {\rm d}\vartheta\int_0^{b}{\rm d}r
     \ e^{-r^2}\frac{r^{m+n+1}}{\sqrt{m!n!}}e^{i(m-n)\vartheta}
     =\frac{2\pi}{ C}\sum_{n=0}^\infty\frac{\ketbra{n}{n}}{ n!}
     \int_0^{b}{\rm d}r\ e^{-r^2}r^{2n+1}\nn\\
    & =\frac{\pi}{ C}\sum_{n=0}^\infty\frac{\ketbra{n}{n}}{ n!}
     \int_0^{b^2}{\rm d}x\ e^{-x}x^n
     =\frac{\pi}{ C}\exp\left(-b^2\right)\sum_{n=0}^\infty
     \left(\exp\left(b^2\right)-\sum_{k=0}^n\frac{b^{2k}}{ k!}\right)\ketbra{n}{n}.
\end{align}
$\ket{\alpha}=D(\alpha)\ket{0}
=\exp(-|\alpha|^2/2)\sum_{n=0}^\infty\frac{\alpha^n}{\sqrt{n!}}\ket{n}$
is a coherent state represented as a displaced vacuum via the
displacement operator $D(\alpha)$ (from now on it will not be
mentioned explicitly that $|\alpha|\leq b$ for all displacement
operators used in this paper and for given $b$) and $C=\pi b^2$ is
a normalization constant. Note that
calculation~(\ref{integral_0b}) for $b\to\infty$ without the
normalization is nothing else than well known resolving of unity
giving the evidence of spanning the whole Hilbert space.

Having defined the maximally mixed state let's investigate which
operations Alice uses for encryption. This transformation should
be as close as possible to the maximally state in the HS~distance
sense and the closeness will depend on the number of used
operations. Beforehand, we will note how to facilitate forthcoming
tedious calculations on a sample example. Suppose that Alice has
e.g. four encryption operations, which displace the vacuum state
to the same distance $r_4$ from the origin but under four
different angles (from symmetrical reason these angles are
multiples of $\frac{\pi}{2}$ in this case). Alice chooses these
operations with the same probability. Now, if we write down the
overall mixture from the states, it can be shown that there exists
a computationally advantageous ``conformation" when the mixture
reads
\begin{equation}\label{mix_on_circle__example}
    \varrho_4=\frac{1}{4}\sum_{q=1}^4\ketbra{\alpha_{4q}}{\alpha_{4q}}
    =\exp(-r_4^2)\sum_{m,n=0}^\infty\frac{r_4^{m+n}}{\sqrt{m!n!}}
    \ketbra{m}{n}\,\delta_m^{m'},
\end{equation}
where $m'=m+4l$, $l=1,\dots,\infty$ and $\delta_m^{m'}=1$ for all
$m$ (occupation number) else $\delta_m^{m'}=0$. Informally,
(\ref{mix_on_circle__example}) is always a real density matrix
with off-diagonal non-zero ``stripes" separated from main diagonal
and from a neighbouring stripe by three zero off-diagonals.
$\varrho_4$ acquires relatively simple
form~(\ref{mix_on_circle__example}) if
$\alpha_{4q}=r_4e^{iq\vartheta_{4q}}$ for
$\vartheta_{4q}=\frac{\pi}{4}+(q-1)\frac{\pi}{2}$. This can be
generalized for arbitrary number of states dispersed on a circle
with given radius $r_p$
\begin{equation}\label{mix_on_circle__general}
    \varrho_p=\frac{1}{ p}\sum_{q=1}^p\ketbra{\alpha_{pq}}{\alpha_{pq}}
    =\exp(-r_p^2)\sum_{m,n=0}^\infty\frac{r_p^{m+n}}{\sqrt{m!n!}}
    \ketbra{m}{n}\,\delta_m^{m'},
\end{equation}
where $p\in\bbZ^+$,$m'=m+pl$ and $\delta_m^{m'}$ is defined as
before. Favorable
``$p$-conformations''~(\ref{mix_on_circle__general}) occur when
$\alpha_{pq}=r_pe^{iq\vartheta_{pq}}$ for
$\vartheta_{pq}=\frac{\pi}{ p}(2q-1)$. Now, we may proceed to the
mixture characterizing all Alice's encryption operations. She
chooses $N\geq1$ and defines $r_p=pb/N$ for $p=1,\dots,N$. Then
\begin{equation}\label{mix_on_circle}
    \Phi_N=\frac{1}{M}\sum_{p=1}^Np\varrho_p
    =\frac{1}{
    M}\sum_{p=1}^N\sum_{q=1}^pD(\alpha_{pq})\ketbra{0}{0}D^\dagger(\alpha_{pq})
\end{equation}
with normalization $M=\frac{N(N+1)}{2}$. To sum up the protocol,
Alice and Bob share a secret and random key which indexes their
operations. So, Alice equiprobably chooses from the set of $M$
displacement operators
$D(\alpha_{pq})=\exp(\alpha_{pq}a^\dagger-\alpha_{pq}^*a)$ where
just one operator creates a coherent state on the circle of radius
$r_1$, two operators generate two states on the circle of radius
$r_2$ and so forth up to $N$. Mixtures of the states on particular
circles are in favourable form~(\ref{mix_on_circle__general}) and
the whole state is~(\ref{mix_on_circle}). The rest of the protocol
is the same as in discrete state encryption. Alice sends the
encrypted state through a quantum channel towards Bob who makes
Alice's inverse operation to decrypt the state.

The keystone in quantum state encryption both discrete and CV is
the fact that an unknown and arbitrary state can be encrypted. If
the state was known we would't need this procedure at all because
Alice could just send information about the preparation of the
state to Bob. So, it would suffice to encrypt this classical
information with the Vernam cipher. Also, our definition of CVPQC
must be independent on the state which is to be encrypted. To
provide this we will find useful unitary invariance of the
HS~distance $D_{HS}(\varrho,\sigma)=D_{HS}(U\varrho
U^\dagger,U\sigma U^\dagger)$ for an arbitrary unitary matrix $U$.
This invariance is, however, necessary but not sufficient
condition for our purpose. The second important issue is due to
advantageous algebraic properties of displacement operators. Let's
demonstrate it on Alice's encryption algorithm which stays the
same as before. If she gets an arbitrary coherent state
$\ket{\beta}$ ($|\beta|\leq b$) she randomly chooses one from
$M$~displacement operators $D(\alpha_{pq})$ (as the shared key
with Bob dictates). Then, even if two displacement operators
generally do not commute we may write a general encryption TPCP
(trace-preserving completely positive) map
\begin{equation}\label{map_Alice}
    \mathcal{E}_N(\ket{\beta})=\frac{1}{
    M}\sum_{p=1}^N\sum_{q=1}^pD(\alpha_{pq})D(\beta)\ketbra{0}{0}D^\dagger(\beta)D^\dagger(\alpha_{pq})
    =\frac{1}{
    M}\sum_{p=1}^N\sum_{q=1}^pD(\beta)D(\alpha_{pq})\ketbra{0}{0}D^\dagger(\alpha_{pq})D^\dagger(\beta)
    =D(\beta)\Phi_ND^\dagger(\beta).
\end{equation}
It is obvious that generally
$\mathcal{E}_N(\ket{\beta})\not=\Phi_N$ (and similarly $\int
D(\alpha)\ketbra{\beta}{\beta}D^\dagger(\alpha){\rm
d}^2\alpha\propto
D(\beta)\openone_bD^\dagger(\beta)=\openone_b^{\beta}\not
=\openone_b$) but their HS~distances are equivalent, i.e.
$D_{HS}(\openone_b^{\beta},\mathcal{E}_N(\ket{\beta}))=D_{HS}(\openone_b,\Phi_N)$.
Thus, we are henceforth entitled to make all calculations of the
HS~distance between $\openone_b^{\beta}$ and
$\mathcal{E}_N(\ket{\beta})$ with explicit
forms~(\ref{integral_0b}) and~(\ref{mix_on_circle}). After some
calculations we will see (Appendix~\ref{app_guess}) that
\begin{equation}\label{HSdist_guess}
    D_{HS}^2\left(\openone_b^{\beta},\mathcal{E}_N(\ket{\beta})\right)
    \approx\left(\frac{1}{
    N+1}\right)^2+\mathcal{O}\left(N^{-4}\right),
\end{equation}
which holds for all for all input coherent states $\ket{\beta}$.
The guess~(\ref{HSdist_guess}) is far from optimal (e.g.
independent on $b$) and is not even an inequality. But this
doesn't matter because the exact form can be derived (for details
see Appendix~\ref{app_exact_value}). Its only problem is relative
complexity so we cannot easily deduce the number of operations for
a given level of secrecy. Nevertheless, (\ref{HSdist_guess})
asymptotically approaches to the exact expression (as is shown in
Appendix~\ref{app_guess}). Notice that in spite of the derivation
of~(\ref{HSdist_guess}) (or, next, analytical
expression~(\ref{HSdist_explicit})) we still cannot reasonably
define a CVPQC. We will do so in section~\ref{sec_threats} after
presenting another assumptions regarding eavesdropping on our
private quantum channel.

If we consider the described unitary invariance of the HS~distance
we write down lhs~of~(\ref{HSdist_guess})
\begin{equation}\label{HSdist_into_traces}
    D_{HS}^2(\openone_b,\Phi_N)
    =\Tr{}\left((\openone_b-\Phi_N)^2\right)
    =\Tr{}\left(\openone_b^2\right)-2\Tr{}\left(\openone_b\Phi_N\right)
    +\Tr{}\left(\Phi_N^2\right).
\end{equation}
After some calculations (see mentioned
Appendix~\ref{app_exact_value}) we get a little bit lengthy
expression
\begin{align}\label{HSdist_explicit}
    D_{HS}^2(\openone_b,\Phi_N)
    & =\frac{1}{ b^2\exp(2b^2)}
    \left(
     \exp(2b^2)-I_0\left(2b^2\right)-I_1\left(2b^2\right)
    \right)
    -\frac{4\exp(-b^2)}{ b^2N(N+1)}
       \sum_{p=1}^N\frac{p}{\exp\left(r_p^2\right)}
       \sum_{k=1}^\infty\left(\frac{b}{ r_p}\right)^kI_k(2r_pb)\nn\\
    & +\left(\frac{2}{ N(N+1)}\right)^2
        \left[\sum_{p=1}^N\frac{p^2}{\exp\left(2r_p^2\right)}
         \left(
          I_0\left(2r_p^2\right)+2\sum_{k=1}^\infty I_{pk}\left(2r_p^2\right)
         \right)
        \right.\nn\\
    & +\sum_{\ahead{p_1,p_2=1}{p_1\not=p_2}}^N\frac{p_1p_2}{\exp\left(r_{p_1}^2+r_{p_2}^2\right)}
        \left.
         \left(
         I_0(2R_{12})+2\sum_{k=1}^\infty I_{p_1p_2k}(2R_{12})
         \right)
        \right],
\end{align}
where $r_p=pb/N,r_{p_l}=p_lb/N,R_{12}=p_1p_2\left(b/N\right)^2$
and $I_n(x)$ is a modified Bessel function of the first kind of
order $n$
\begin{equation}\label{bessfce}
    I_n(x)=\sum_{s=0}^\infty{\left(\frac{x}{2}\right)^{n+2s}}\frac{1}{(n+s)!s!}.
\end{equation}
However, as we declared, for a rough estimation of the number of
operations for a given secrecy we will
employ~(\ref{HSdist_guess}). Using
$M=\frac{N(N+1)}{2}\approx\frac{(N+1)^2}{2}$ (and then again
$M\approx\frac{N^2}{2}$) we find that the dependence of the number
of bits on the HS~distance~(\ref{HSdist_explicit}) is $n=\log
M=-2(1+\log D_{HS})+\log \left(1+\sqrt{1+CD_{HS}^2}\right)
\approx-1-2\log D_{HS}$ because the guess is in particular precise
for $D_{HS}\ll1$. $C$ is a constant (polynomial) bounding $N^{-4}$
in~(\ref{HSdist_guess}) from above. By the numerical simulations
based on~(\ref{HSdist_explicit}) for different $b$ and $M$ we see
that the guess is accurate and for higher $M$ even serves as a
relatively good upper bound.

\subsection*{Simplified encryption}

Now, let's see what is going to happen if we simplify our
encryption protocol. The arrangement is the same, i.e. Alice's
task is to encipher single-mode coherent states and her only
knowledge is that the states are somewhere inside the circle of
radius $r\leq b$. However, Alice's technology is limited and we
will tend to replace technologically demanding displacements by
simple phase shifts given by the well known time evolution
$\ket{\alpha(t)}=\exp(-iHt/\hbar)\ket{\alpha}$ with the
Hamiltonian $H=\hbar\omega(n+\frac{1}{2})$. The resulting state is
$\ket{\alpha(t)}=\exp(-i\omega t/2)\ket{\alpha\exp(-i\omega t)}$
so the state undertakes the rotation about $\Theta=\omega t$
regardless the distance (which stays preserved) from the phase
space origin. Note that unitary operator $U=\exp(-iHt/\hbar)$ has
the form
\begin{equation}\label{unitary_evolution}
    U=\exp\left(-\frac{i\omega t}{2}\right)\sum_{n=0}^\infty\exp(-i\omega t n)
    \ketbra{n}{n}.
\end{equation}
Next, suppose that Alice is equipped with $p$ encryption
operations where she can rotate the state about $q$~multiples of
$2\pi/p$. So, $\Theta_q=q2\pi/p$ where $q=1,\dots,p$. Then, for
someone without any information about the angle of rotation chosen
by Alice (e.g. Eve) the state leaving Alice is in the form
\begin{equation}\label{ro_p_tilde}
\tilde\varrho_p=\frac{1}{
p}\sum_{q=1}^{p}U_q\ketbra{\alpha}{\alpha}U^{-1}_q=\frac{1}{
p}\sum_{q=1}^{p}\left(\ketbra{\alpha}{\alpha}\right)_q=\frac{1}{
p}\sum_{q=1}^{p}\ketbra{\alpha e^\frac{q2\pi}{p}}{\alpha
e^\frac{q2\pi}{ p}}
\end{equation}
where $U_q$ is~(\ref{unitary_evolution}) with $\Theta_q=\omega
t_q=q2\pi/p$. Because Alice doesn't know where exactly the state
$\ket{\alpha}$ is placed we cannot write an explicit form of
$\tilde\varrho_q$. Apparently, however, for arbitrary $p$ it can
be transformed to our favourite
state~(\ref{mix_on_circle__general}), again with the help of
unitary operation~(\ref{unitary_evolution})
\begin{equation}\label{ro_p}
   \varrho_p(r_p)=U\tilde\varrho_pU^{-1}
   =\exp(-r_p^2)\sum_{m,n=0}^\infty\frac{r_p^{m+n}}{\sqrt{m!n!}}
    \ketbra{m}{n}\,\delta_m^{m'},
\end{equation}
where $m'=m+pl$, $l=1,\dots,\infty$. It remains to show that
$\openone_b=U\openone_bU^{-1}$. This can be readily seen from the
fact that~(\ref{unitary_evolution}) is diagonal. This is vital.
Now we can easily calculate the HS~distance of $\varrho_p(r_p)$
and $\openone_b$ and regarding the unitary invariance of the
HS~distance the result will be valid for any arbitrary input
coherent state (this trick is akin to that one used
in~(\ref{map_Alice})). After the calculation we get the same
result as in~(\ref{HSdist_explicit}) but for fixed $p$ and without
the summation over $p_1,p_2$
\begin{align}\label{HSdist_simple_explicit}
    D_{HS}^2(\openone_b,\varrho_p)
    & =\frac{1}{ b^2\exp(2b^2)}
     \left(
      \exp(2b^2)-I_0\left(2b^2\right)-I_1\left(2b^2\right)
     \right)
     -\frac{2}{\exp\left(b^2\right)b^2}
       \frac{1}{\exp\left(r_p^2\right)}
       \sum_{k=1}^\infty\left(\frac{b}{ r_p}\right)^kI_k(2r_pb)\nn\\
    & +\frac{1}{\exp\left(2r_p^2\right)}
         \left(
          I_0\left(2r_p^2\right)+2\sum_{k=1}^\infty I_{pk}\left(2r_p^2\right)
         \right).
\end{align}
\begin{figure}[t]
\resizebox{16.5cm}{6.4cm}{\includegraphics{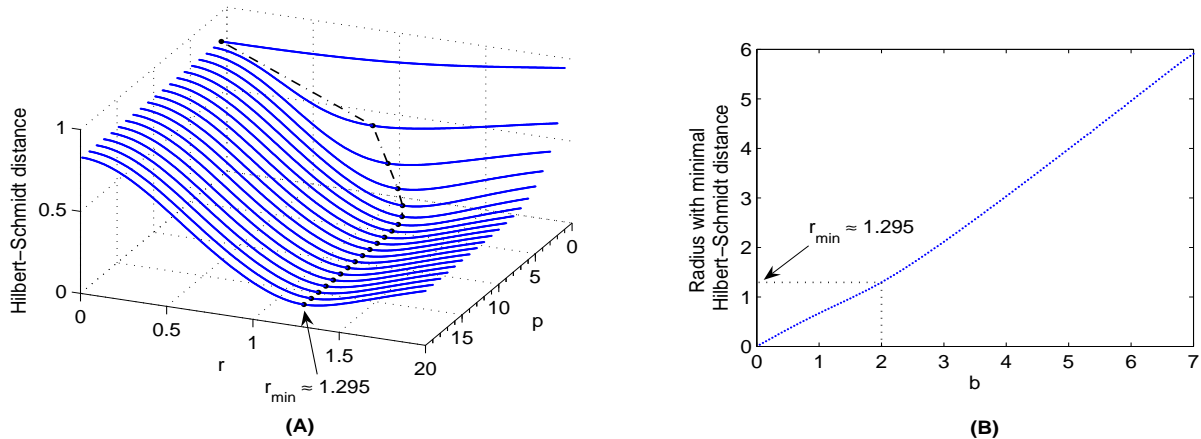}}
\caption{\label{figs_saturated_min}(A) In the case of
technologically simplified encryption Alice has at disposal just
one displacement operator and several phase shifts. The depicted
HS~distances~(\ref{HSdist_simple_explicit}) (here for $b=2$ and
$p=1,\dots,20$ determining the number of used encryption
operations) illustrates how Alice chooses the proper
displacement~$D(r_{min})$ and a correct number of phase shifts. We
see that for $p\agt6$ the minimum is ``saturated'' so Alice needs
just six phase shifts for the best approximation to
full-encryption protocol. Another rising of~$p$ doesn't bring any
advantage. The dash-dotted curve connects minima for their better
visibility.~(B)~Numerical solutions of~Eq.~(\ref{min_D_HS}) for
different~$b$ (i.e. growing input range of states coming to Alice)
tells us which displacement is needed in the case of simplified
encryption. For these values of $r_{min}$ the
HS~distance~(\ref{HSdist_simple_explicit}) reaches the lowest
values. A special case of $b=2$ is highlighted to see its
correspondence with the picture on the left.}
\end{figure}
The behaviour of this quantity is completely different compared
to~(\ref{HSdist_explicit}) describing the original case as can be
seen in Fig.~\ref{figs_saturated_min}~(A). As an example we
put~$b=2$ and we see that there exists $r_{min}<b$ for which the
HS~distance between~(\ref{integral_0b}) and~(\ref{ro_p}) reaches
its minimum for sufficiently high~$p$. In order to find a radius
minimizing the HS~distance of these density matrices for
various~$b$ (we put $p\to\infty$ and thus $2\sum_{k=1}^\infty
I_{pk}\left(2r_p^2\right)$ vanishes -- we seek for, say, a
saturated minimum which is independent on the number of operations
on a given circle) we can calculate
\begin{equation}\label{min_D_HS}
    \frac{\partial\left(D_{HS}^2(\openone_b,\varrho_p)\right)}{\partial r}
    \stackrel{p\to\infty}{=}\frac{4}{\exp\left(2r_{min}^2\right)}
    \left(
      r_{min}I_1\left(2r_{min}^2\right)-r_{min}I_0\left(2r_{min}^2\right)
      +\frac{\exp\left(2r_{min}^2\right)}{b\exp\left(b^2\right)}I_1\left(2r_{min}b\right)
    \right)=0.
\end{equation}
The solution is the root of the expression in parentheses.
Unfortunately, an analytical solution wasn't found so numerical
calculation of $r_{min}$ for several $b\in\left(0,7\right.\rangle$
is depicted in Fig.~\ref{figs_saturated_min}~(B).

What are the physical consequences of this whole simplification?
From Fig.~\ref{figs_saturated_min}~(A) we see that a maximally
mixed state $\openone_b$ can be in the distance sense substituted
with a mixture of coherent state on a circle (with radius
$r_{min}$ where the diversion from $\openone_b$ is smallest) and
the mixing requires relatively few encryption operations (low $p$
for the distance saturation). It holds not only for $b=2$ what is
the case in the picture. So, now we can make the process of
encryption/decryption technologically easier. We equip Alice/Bob
with mentioned phase shifts leading to~(\ref{ro_p}) and just one
displacement operator $D(r_{min})$. Every incoming coherent state
is first displaced and then encrypted by choosing a phase shift
(indexed by a secret key). The role of
Fig.~\ref{figs_saturated_min}~(B) is in a proper choice of
$r_{min}$ for given~$b$. Finally, Alice examines
Fig.~\ref{figs_saturated_min}~(A) to select a sufficient
number~$p$ of phase shifts for which the minimum
of~HS~distance~(\ref{HSdist_simple_explicit}) stays essentially
the same.

The important point is that incoming coherent states come to Alice
randomly and equiprobably for given $b$ so after this simplified
encryption (and for many encryption instances) they are as close
as possible to full-fashioned encrypted states as they would be if
processed by using~(\ref{mix_on_circle}), i.e. the case of fully
technologically equipped Alice and Bob, respectively. The overall
advantage is the use of just one technologically demanding
operation which doesn't need to be tuned for given $b$ (it stays
fixed in the encryption protocol described at the beginning of
chapter~\ref{sec_main}).

\section{Eavesdropping on encrypted CV states}
\label{sec_threats}

Quantum channels are TPCP maps on quantum states. Following the
classical channel coding theorem~\cite{shannon48} one may ask how
much information is a quantum channel able to convey. This
question naturally leads to the definition of quantum channel
capacity (for a nice survey on various quantum channel capacities
see Ref.~\cite{shor_capa}) as the maximum of the accessible
information over the probability distributions of the input states
ensemble $\{p_i,\sigma_i\}$ entering the channel (after
classical-quantum coding onto input quantum states
$x_i\rightleftharpoons\sigma_i$ the ensemble of classical messages
is indicated by the variable $X=\{p(x_i),x_i\}$ where
$p(x_i)\equiv p_i$). The accessible information $I_{acc}$ itself
is the maximization over all measurement of the mutual information
$I(X;Y)$ with the variable $Y=\{p\,(y_j),y_j\}$ giving the
probability $p\,(y_j)$ of the result~$y_j$ (an output alphabet) of
a measurement on the channel output. Sometimes it might be
difficult to calculate the accessible information to determine the
channel capacity. Then, we can estimate the accessible information
from above by the Holevo bound
\begin{equation}\label{holevo_bound}
    \chi\left(\{p_i,\sigma_i\}\right)
    =S\biggl(\sum_ip_i\sigma_i\biggr)-\sum_ip_iS(\sigma_i),
\end{equation}
for which was proved~\cite{holevo_bound} that $\chi\geq I_{acc}$
($S(\varrho)$ is the von~Neumann entropy) reaching equality if all
$\sigma_i$ commute.

In our case we try to get as close as possible to our maximally
mixed state~(\ref{integral_0b}) with TPCP
map~(\ref{mix_on_circle}) (or, generally, to~$\openone_b^\beta$
with~(\ref{map_Alice})) which both belong to the class of bosonic
channels~\cite{bos_chan1,bos_chan2}. However, our task is quite
different from reaching the highest quantum channel capacity (what
means ``tuning'' of $p_i$). Now, the variable $X$ is fixed. For
our purpose the Holevo bound can tell us what is the upper bound
on information which is a third party (Eve) able to learn from her
very best (that is optimal) measurement on the quantum channel.
The question now is of what state we should calculate the Holevo
bound to find Eve's maximum of attainable information on {\em
incoming} states. Unlike the discrete case (perfect and
approximate encryption) the condition of perfect and acceptable
closeness of an encrypted state to the maximally mixed state is
not in this case sufficient. Let's demonstrate the reason for this
difference on the Bloch sphere (i.e. perfect qubit encryption). If
Alice encrypts an unknown qubit (with a PQC compound from e.g.
four Pauli matrices) she gets a maximally mixed two-qubit state
(normalized unity matrix). It means that Eve cannot construct any
measurement giving her information on the state. Moreover, even if
she had a priori information on the state (in the sense that Alice
gets and encrypts many copies of the same unknown state) she
wouldn't be able to use any method of unknown quantum states
reconstruction~\cite{quantstateest}. She wouldn't get any clue
where to find the state because all are transformed (encrypted) to
the maximally mixed state dwelling in the center of the sphere.
Unfortunately, this is not our case. Here, if Alice encrypts many
identical coherent states (howbeit unknown) Eve could in principle
reconstruct in which part of the phase space the encryption had
been carried out. Because we next suppose that encryption
operations are publicly known (naturally not the secret key
sequence itself) Eve then could be able to deduce the original
state from many variant ciphers of the same state. So we will
suppose that incoming state do not exhibit such statistics and,
moreover, they are distributed equiprobably and randomly in the
whole region of considering (i.e. within the circle of radius
$r\leq b$)~\footnote{To avoid misunderstanding we distinguish in
this chapter between distribution and statistics. As usually,
distribution means a probability of occurrence for incoming states
or, eventually, for encryption operations. By contrast, statistics
means correlation or better relationship between incoming states
what may be publicly known, e.g. information that several incoming
states will be the same. This has no influence on the
distribution.}. Less restrictive requirement could be a
distribution independent on the phase and changing only with the
distance from the origin (rotationally invariant) what will be
discussed at the end of this chapter.

Based on theses thoughts, we may finally define CVPQC. We call the
object
$\{\mathcal{B},P_{\mathcal{B}},\mathcal{E}_N(\ket{\beta}),\openone_b^{\beta}\}$
as CVPQC where $\mathcal{B}$ is the set of all coherent states
$\ket{\beta}$ ($|\beta|\leq b$) with distribution
$P_{\mathcal{B}}$ around the origin of phase space,
$\mathcal{E}_N(\ket{\beta})$ is the TPCP~map defined
in~(\ref{map_Alice}) and $\openone_b^{\beta}$ is maximally mixed
state centered around an input state $\ket{\beta}$
(displaced~(\ref{integral_0b})). Let's note that the definition of
CVPQC is valid for all coherent mixed states of the general form
$\varrho_\beta=\sum_ip_i(\ketbra{\beta}{\beta})_i$ with
$\sum_ip_i=1$ and $|\beta|_i\leq b,\,\forall i$ because of
convexity of~HS~norm. The situation is similar as
in~\cite{approx_encryption} for the operator norm.

Considering the assumptions from the previous paragraph let's
focus on the problem of security on the channel. For our purpose
we employ an integral version of~(\ref{holevo_bound}). First note
that we are now interested in the limiting case when Alice uses
infinitely many encryption operations~\footnote{Of course, in
reality for every $\ket{\beta}$ Alice produces a ``finite''
mixture $\mathcal{E}_N(\ket{\beta})$ from~(\ref{map_Alice}) which
should be as close as possible to
$\openone_b^\beta=D(\beta)\openone_bD^\dagger(\beta)$. For the
examination how these states are close to each other and its
connection to the number of bits of the key we have derived
expression~(\ref{HSdist_guess}).}. More importantly, it remains to
correctly answer the question posed above. That is, what actually
calculate as the Holevo bound? We know that Eve will measure on
the encrypted channel to get any information on incoming states
(states before the encryption). To learn something about position
of a coherent state before encryption Eve must try to distinguish
among objects which preserve some information about input states
location in phase space. In other words Eve has to choose a
correct ``encoding'' $\sigma_i$ from Eq.~(\ref{holevo_bound}) but
now for a continuous index. Due to the used encryption
operation~(\ref{map_Alice}) (which in our case $N\to\infty$
transforms to $\openone_b^\beta$ as is proved
in~Appendix~\ref{app_exact_value}) a state $\openone_b^\beta$ is
that object which preserves the information on placement of an
incoming state~$\ket{\beta}$ because is centered around it. Then,
a continuous version of~(\ref{holevo_bound}) reads
\begin{equation}\label{holevo_bound_int1}
    \chi\bigl(\{P(\beta),\openone_b^\beta\}\bigr)
    =S\bigl(\tilde\Lambda_{2b}\bigr)
    -\int{\rm
    d^2}\beta\,P(\beta)S\bigl(\openone_b^\beta\bigr),
\end{equation}
where $\tilde\Lambda_{2b}$ is a total state leaving Alice's
apparatus (normalized Eq.~(\ref{Lambda})), $P(\beta)$ is an input
distribution from the CVPQC definition and
$\openone_b^\beta=D(\beta)\openone_bD^\dagger(\beta)$. Since
$S(D(\beta)\openone_bD^\dagger(\beta))=S(\openone_b)$ for all
$\beta$ it immediately follows from~(\ref{holevo_bound_int1})
\begin{equation}\label{holevo_bound_int2}
    \chi\left(\{P(\beta),\openone_b^\beta\}\right)
    =S\bigl(\tilde\Lambda_{2b}\bigr)-S\bigl(\openone_b\bigr).
\end{equation}
\begin{figure}[t]
\resizebox{10cm}{7.5cm}{\includegraphics{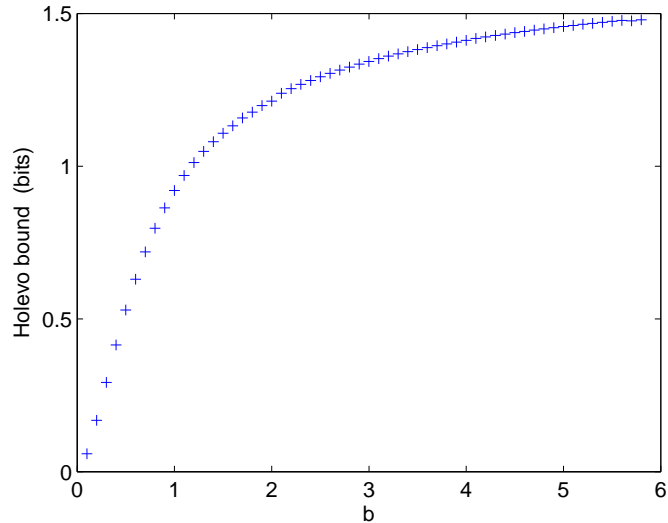}}
\caption{\label{fig_holevo_bound}The Holevo bound represents an
upper bound for information Eve is able to get from her best
measurement on the encrypted channel. The bigger is the set of
states coming to Alice (growing $b$) the more information Eve can
extract. This is caused by better discriminability of Eve's
measuring apparatus. Nevertheless, relatively low calculated
values of the Holevo bound for our CVPQC given by
Eq.~(\ref{holevo_bound_int2}) limits Eve in eavesdropping on the
channel.}
\end{figure}
To calculate the Holevo bound~(\ref{holevo_bound_int2}) it remains
to find the total state after the encryption procedure provided
that Eve doesn't know neither which particular coherent
state~$\ket{\beta}$ comes to Alice nor the key used for
encryption. Thus, because the incoming distribution
$P(\beta)\equiv P(x,\vartheta)=1$ (no need to normalize it here)
is publicly known then Alice quite accidentally experiences
$\openone_b$ also as a total {\em incoming} state. So, the state
{\em leaving} Alice is in the form
\begin{align}\label{Lambda}
\Lambda_{2b}
& = \int{\rm d^2}\alpha D(\alpha)
  \left(\int{\rm d^2}\beta\,\ketbra{\beta}{\beta}\right)D^\dagger(\alpha)
=\int_{0}^{b}\int_{0}^{2\pi}{\rm d^2}\beta
 \left(
  \int_{0}^{b}\int_{0}^{2\pi}{\rm d^2}\alpha\,
  D(\alpha)\ketbra{\beta}{\beta}D^\dagger(\alpha)
 \right)\nn\\
& =\int_{0}^{b}x{\rm d}x\int_{0}^{b}y{\rm d}y\int_{0}^{2\pi}{\rm
  d}\varphi
 \left(
  \int_{0}^{2\pi}{\rm d}\vartheta\,\ketbra{\beta+\alpha}{\beta+\alpha}
 \right)
=2\pi\sum_{n=0}^{\infty}\frac{\ketbra{n}{n}}{n!}
   \int_{0}^{b}\int_{0}^{b}\int_{0}^{2\pi}{\rm d}x{\rm d}y{\rm d}\varphi
   \ e^{-R^2}R^{2n+1}xy,
\end{align}
where $R^2=x^2+y^2-2xy\cos\varphi\leq4b^2$, ${\rm d^2}\alpha=y{\rm
d}y{\rm d}\varphi$, ${\rm d^2}\beta=x{\rm d}x{\rm d}\vartheta$ and
$D(\alpha)$ are encryption displacements (again uniformly
distributed as the CVPQC definition dictates and without a
normalization). Shortly said, by changing the order of integration
we see that $\Lambda_{2b}$ is diagonal. Concretely, the term in
the parentheses in the reordered integral (second row
in~(\ref{Lambda})) is an ordinary coherent state
$\ket{\beta+\alpha}$ ($r$ is its distance from the origin and
abscissae formed by points in the phase space $(\alpha,\beta)$ and
$(\beta,0)$ contain a relative angle $\varphi$) and is integrated
out over the phase. This gives a diagonal matrix and next
integrating just changes the statistics on the diagonal. Note that
$\Lambda_{2b}$ ``covers'' the area of radius~$2b$ in phase space.
Unfortunately, it is difficult to obtain an analytical solution
of~(\ref{Lambda}). Nevertheless, based on~(\ref{Lambda}) we may
generally write (tilde indicates normalization)
$\tilde\Lambda_{2b}=\sum_n\lambda^{(b)}_n\ketbra{n}{n}$, where
$\lambda^{(b)}_n$ are needed to be determined numerically.
Inserting $\tilde\Lambda_{2b}=\sum_n\lambda^{(b)}_n\ketbra{n}{n}$
and~(\ref{integral_0b}) into Eq.~(\ref{holevo_bound_int2}) we find
the desired Holevo bound what is depicted in
Fig.~\ref{fig_holevo_bound}. It is interesting to note that the
convergence of the Holevo bound is tightly connected with the
energy constraint $\int\Tr{}\left(\varrho_\beta a^\dagger
a\right){\rm d}P_{\mathcal{B}}\leq\bar n_b$ which is automatically
satisfied through the finiteness of~$b$ in our CVPQC. For a
general distribution~$P(\beta)$ it is satisfied due to the natural
condition $\int{\rm d^2}\beta P(\beta)=1$. The importance of the
constraint has been already recognized for capacities of bosonic
channels~\cite{bos_chan1,bos_chan2,bos_chan_lossy,bos_gauss_chan_mem}.

From Fig.~\ref{fig_holevo_bound} we see that there is no chance
for Eve to find which coherent state actually passes on the
channel. This is not so surprising considering the non-discrete
character of the input distribution $P(\beta)$. But Eve cannot
divide the appropriate area at least roughly to approximately
devise the position of a particular encrypted state
$\openone_b^\beta$ (and from this to derive the desired position
of an incoming state $\ket{\beta}$). From the informational point
of view the Holevo bound results could be interpreted such that
for a given~$b$ there doesn't exist any optimal measurement
enabling Eve to divide the circle of the radius~$2b$ in more
than~$2^{\chi}$ sections and tells her in which one the encrypted
state occurs.

Having fixed the input distribution $P(\beta)$ of coherent states
coming to Alice another way how to next decrease the Holevo bound
might be a different definition of an encryption distribution
in~Eq.~(\ref{Lambda}) where a uniform distribution is implicitly
used. If this distribution was symmetrical (i.e. rotationally
invariant, for example Gaussian one) then we would
find~Eq.~(\ref{Lambda}) diagonal as well and thus we could easily
calculate~(\ref{holevo_bound_int2}). However, a big challenge is
proving the optimality of such distribution function. Anyhow, it
means that by minimizing the Holevo
bound~(\ref{holevo_bound_int2}) within the context of the CVPQC
definition we could tackle the previously mentioned problem of
maximality of our mixed state by putting this more suitable
distribution into Eq.~(\ref{integral_0b}).

\section{Conclusion}

In this work we have opened the problem of continuous variable
encryption of unknown quantum states and introduced the concept of
private quantum channels (CVPQC) into this area. For the start we
have restricted ourselves on coherent states which belong to the
important class of states with the Gaussian distribution function.
A~particular continuous variable private quantum channel was
proposed and we have studied its properties. Firstly, it means
that we have established the notion of so called maximally mixed
state with regard to its non-discrete (continuous variable)
nature. For this kind of mixture we were interested how many
encryption operations are sufficient to consider an incoming
coherent state to be secure. This quantity was determined by
calculating the Hilbert-Schmidt distance between the mixture and
the encrypted state for an arbitrary number of encryption
operations. Next, we have studied the possibility of eavesdropping
on the quantum channel. We have supposed that Eve is able to
perform an optimal measurement to get the maximum information on
the state (which is limited by the calculated Holevo bound) or is
able to use some quantum state estimation methods. The second
possibility (which requires Eve's a priori information on the
statistics of the states in the sense that she knows that Alice
gets many copies of the same unknown coherent state to encrypt)
restricts the CVPQC definition. The way how to avoid this kind of
attack is the most desired direction of next research.

Beside the above mentioned topic we have addressed many more
intriguing questions which can stimulate another research in this
area and improve the existing protocol. Among many topics let's
name the problem of either universal distribution of incoming
state or a universal distribution of encryption operations or
both. This is tightly connected with the freedom in choice of the
definition of maximally mixed state. We have seen that there
exists a certain ambiguity in the definition of what we call here
the maximally mixed state in phase space. Its relevance is
measured by the accessible information (or eventually limited from
above by the Holevo bound) Eve can get and the question is what
kind of definition of the maximally mixed state is the most
appropriate for a given incoming distribution of states. Together
with this topics another generalization presents itself. It is the
possibility to encrypt another Gaussian states, especially
squeezed states.

\begin{acknowledgments}
The author is grateful to L.~Mi\v sta, R.~Filip and J.~Fiur\'a\v
sek for useful discussions in early stage of this work, M.~Du\v
sek for reading the manuscript and T.~Hol\'y for granting the
computational capacity. The support from the EC project SECOQC
(IST-2002-506813) is acknowledged.
\end{acknowledgments}

\appendix
\section{}
\label{app_exact_value}

Eq.~(\ref{HSdist_into_traces}) consists of three parts. Let's
calculate them step by step. If we rewrite~(\ref{integral_0b}) in
a more suitable form we get diagonal elements in the form
\begin{equation}\label{app_integral0b}
    \openone_b(n,n)=\frac{1}{b^2\exp(b^2)}
    \left(\sum_{m=n+1}^\infty\frac{b^{2m}}{m!}\right)\ketbra{n}{n}.
\end{equation}
Now, we may easily calculate
\begin{align}
  \Tr{}\left(\openone_b^2\right)
    & = \frac{1}{b^4\exp(2b^2)}
     \sum_{n=1}^\infty\left(\sum_{m=n}^\infty\frac{b^{2m}}{ m!}\right)^2
     = \frac{1}{b^4\exp(2b^2)}\left(\sum_{n=1}^\infty\frac{b^{4n}}{(n!)^2}n
     + 2\sum_{k=1}^\infty\sum_{n=1}^\infty\frac{b^{4n+2k}}{ n!(n+k)!}n\right)\nn  \\
    & = \frac{1}{b^4\exp(2b^2)}\left(\sum_{n=0}^\infty\frac{b^{4(n+1)}}{ n!(n+1)!}
     + 2\sum_{k=1}^\infty\sum_{n=0}^\infty\frac{b^{4(n+1)+2k}}{n!(n+k+1)!}\right)
     = \frac{1}{b^2\exp(2b^2)}\left(I_1\left(2b^2\right)
     + 2\sum_{k=1}^\infty I_{k+1}\left(2b^2\right)\right)\nn\\
    & =
    \frac{1}{b^2\exp(2b^2)}\left(\exp(2b^2)-I_0\left(2b^2\right)-I_1\left(2b^2\right)\right),
\end{align}
where $I_s$ is a modified Bessel function of the first kind of
order $s$. The last row is due to identity
\begin{equation}\label{bess_ident}
    I_0(x)+2\sum_{k=1}^\infty I_k(x)=\exp(x).
\end{equation}
For the calculation of the cross element
$\Tr{}\left(\openone_b\Phi_N\right)$ we use~(\ref{app_integral0b})
again. But for the sake of clarity we will calculate only
$p\Tr{}\left(\openone_b\varrho_p\right)$. The overall trace is
given by summation $\Tr{}\left(\openone_b\Phi_N\right)=1/M
\sum_{p=1}^Np\Tr{}\left(\openone_b\varrho_p\right)$ what follows
from~(\ref{mix_on_circle}). From~(\ref{mix_on_circle__general}) we
get a general diagonal element
\begin{equation}\label{prhop}
    p\varrho_p(n,n)=p\exp\left(-r_p^2\right)\frac{r_p^{2n}}{ n!}
    \ketbra{n}{n}
\end{equation}
and then
\begin{align}
  p\Tr{}\left(\openone_b\varrho_p\right)
   &=  \frac{1}{ b^2\exp(b^2)}\frac{p}{\exp(r_p^2)}
     \sum_{n=0}^\infty\sum_{m=n+1}^\infty\frac{r_p^{2n}}{ n!}\frac{b^{2m}}{ m!}
    = \frac{1}{ b^2\exp(b^2)}\frac{p}{\exp(r_p^2)}
     \sum_{k=1}^\infty\sum_{n=0}^\infty\frac{b^{2(n+k)}}{ n!(n+k)!}r_p^{2n}\nn \\
   &= \frac{1}{ b^2\exp(b^2)}\frac{p}{\exp(r_p^2)}\sum_{k=1}^\infty
     \left(\frac{b}{ r_p}\right)^kI_k(2r_pb).
\end{align}
Overall, we have
\begin{equation}\label{cross}
    \Tr{}\left(\openone_b\Phi_N\right)
    =\frac{2}{ N(N+1)}\frac{1}{ b^2\exp(b^2)}
    \sum_{p=1}^Np\exp\left(-r_p^2\right)\sum_{k=1}^\infty
     \left(\frac{b}{ r_p}\right)^kI_k(2r_pb).
\end{equation}
The third part can be written
\begin{equation}\label{Phi_n^2}
    \Tr{}\left(\Phi_N^2\right)=\left(\frac{2}{N(N+1)}\right)^2
    \Tr{}\left(\sum_{i,j=1}^{N}p_ip_j\varrho_{p_{i}}\varrho_{p_{j}}\right)
\end{equation}
where particular summands have the form
\begin{equation}\label{rhoirhoj}
   \Tr{}\left(p_ip_j\varrho_{p_{i}}\varrho_{p_{j}}\right)
   =\frac{p_ip_j}{\exp\left(r_{p_i}^2+r_{p_j}^2\right)}
   \left(
         I_0(2R_{ij})+2\sum_{k=1}^\infty I_{p_ip_jk}(2R_{ij})
   \right),
\end{equation}
what can be easily seen if we
substitute~(\ref{mix_on_circle__general}) into a slight
generalization of~(\ref{bess_ident})
\begin{equation}\label{bess_ident2}
   I_0(2xy)+2\sum_{k=1}^\infty I_k(2xy)=\exp(x^2)\exp(y^2),
\end{equation}
where for our purpose
$x=r_{p_i}=p_ib/N,y=r_{p_j}=p_jb/N,R_{ij}=p_ip_j(b/N)^2$.

\section{}
\label{app_guess}

When performing a limit transition it can be shown that
$\lim_{N\to\infty}|\Phi_N-\openone_b|$=0. The only nonzero
elements of $\Phi_N$ stays on its diagonal and are equal to
diagonal elements of $\openone_b$. Let's demonstrate it on first
diagonal element ($n=0$). From~(\ref{mix_on_circle__general})
and~(\ref{mix_on_circle}) we see that
\begin{equation}\label{Phi_N00}
    \Phi_N(0,0)=\frac{2}{ N(N+1)}\sum_{p=1}^Np\exp(-r_p^2)\ketbra{0}{0}
    =\frac{2}{ N(N+1)}\sum_{p=1}^N\sum_{n=0}^\infty(-1)^n{b^{2n}}{ n!}
    \frac{p^{2n+1}}{ N^{2n}}\ketbra{0}{0}.
\end{equation}
Assuming that
\begin{equation}\label{powersum}
    \sum_{p=1}^Np^d=\frac{1}{
    d+1}\sum_{p=1}^{d+1}(-1)^{d+1-p}\binom{d+1}{p}B_{d+1-p}N^p,
\end{equation}
where $B_n$ are Bernoulli numbers, inserting the highest
polynomial $\frac{N^{d+1}}{ d+1}$ (lower polynomials subsequently
tends to zero) from~(\ref{powersum}) into~(\ref{Phi_N00}) and
finally letting $N\to\infty$ we get (omitting ketbra)
\begin{equation}\label{PhiN00_trans}
   \lim_{N\to\infty}\Phi_N(0,0)=2\lim_{N\to\infty}\sum_{n=0}^\infty
   (-1)^n\frac{b^{2n}}{ n!}\frac{N^{2(n+2)}}{ N^{2n+1}(N+1)}\frac{1}{2(n+1)}
   =\sum_{n=1}^\infty(-1)^n\frac{b^{2n}}{ n!}={1-\exp(-b^2)}{
   b^2}\equiv\openone_b(0,0)
\end{equation}
as can be seen from~(\ref{app_integral0b}). In this way we could
continue for all diagonal elements of $\Phi_N(n,n)$. But turn our
attention elsewhere. If we realize that a general diagonal element
is of the form
\begin{equation}\label{PhiNnn}
    \Phi_N(n,n)=\frac{2}{ N(N+1)}\frac{1}{n!}
    \sum_{p=1}^Np\exp(-r_p^2)r_p^{2n}\ketbra{n}{n},
\end{equation}
we see that the expression for the
HS~distance~(\ref{HSdist_explicit}) can be simplified in the
following way
\begin{align}\label{HS_simple}
   D_{HS}^2(\openone_b,\Phi_N)& =|\openone_b-\Phi_N|^2\\
   & +
   2\left(\frac{2}{N(N+1)}\right)^2
        \left[\sum_{k=1}^N\frac{k^2}{\exp\left(2r_k^2\right)}
         \left(
          \sum_{n=1}^\infty I_{kn}\left(2r_k^2\right)
         \right)
         +\sum_{\ahead{k_1,k_2=1}{k_1\not=k_2}}^N\frac{k_1k_2}{\exp\left(r_{k_1}^2+r_{k_2}^2\right)}
         \left(
          \sum_{n=1}^\infty I_{k_1k_2n}(2R_{12})
         \right)
        \right].\nn
\end{align}
This form will help us in deriving~(\ref{HSdist_guess})
\begin{equation}\label{guess}
    |\openone_b-\Phi_N|=\sum_{k=0}^\infty|\openone_b(k,k)-\Phi_N(k,k)|
    \approx\sum_{k=0}^\infty\left|\openone_b(k,k)\left(1-\frac{N}{ N+1}\right)\right|
    =\frac{1}{ N+1}
\end{equation}
and hence
$|\openone_b-\Phi_N|^2\approx\left(\frac{1}{N+1}\right)^2$. We
again put the highest polynomial $\frac{N^{d+1}}{d+1}$ into
$\Phi_N(k,k)$ and used the normalization condition
$\sum_{k=0}^\infty\openone_b(k,k)=1$. Second part
of~(\ref{HS_simple}) tends to zero even faster. We can see it with
the help of~(\ref{bess_ident2}) from which follows that
$\sum_{n=1}^\infty
I_{kn}\left(2r_k^2\right)\ll\exp\left(2r_k^2\right)$ especially
for higher $k$ (and, of course, similarly for the second summand
in square brackets of expression~(\ref{HS_simple})). From these
considerations~(\ref{HSdist_guess}) follows and, not surprisingly,
asymptotically approaches the exact value.

\end{document}